\documentclass[conference]{IEEEtran}
\IEEEoverridecommandlockouts
\usepackage{latexsym}
\usepackage{graphicx}
\usepackage{amsfonts,amssymb,amsmath}
\def\BibTeX{{\rm B\kern-.05em{\sc i\kern-.025em b}\kern-.08em T\kern-.1667em\lower.7ex\hbox{E}\kern-.125emX}}

\usepackage{algorithmic}
\usepackage{algorithm}

\usepackage{subfig}
\usepackage{siunitx}

\usepackage[acronym]{glossaries}
\newcommand{\newac}{\newacronym}
\newcommand{\ac}{\gls}

\makeglossaries

\newac{los}{\text{LoS}}{line-of-sight}
\newac{nlos}{\text{NLoS}}{non-line-of-sight}

\DeclareMathAlphabet{\mathbit}{OML}{cmr}{bx}{it}
\DeclareMathAlphabet{\mathsf}{OT1}{cmss}{m}{n}
\DeclareMathAlphabet{\mathTXf}{OT1}{cmss}{bx}{it}

\DeclareMathOperator{\Transpose}{T}

\DeclareMathOperator{\Exp}{E}

\DeclareMathAlphabet{\mathpzc}{OT1}{pzc}{m}{it}

\newcommand{\Tr}{{\Transpose}}

\usepackage{anyfontsize}

\begin{document}

\title{Model-aided Deep Reinforcement Learning for Sample-efficient UAV Trajectory Design in IoT Networks\\
}

\author{
\IEEEauthorblockN{Omid Esrafilian, Harald Bayerlein, and David Gesbert}
\IEEEauthorblockA{Communication Systems Department, EURECOM, Sophia Antipolis, France\\
\text{\{omid.esrafilian, harald.bayerlein, david.gesbert\}@eurecom.fr}
}
}

\maketitle

\begin{abstract} Deep Reinforcement Learning (DRL) is gaining attention as a potential approach to design trajectories for autonomous unmanned aerial vehicles (UAV) used as flying access points in the context of cellular or Internet of Things (IoT) connectivity. DRL solutions offer the advantage of on-the-go learning hence relying on very little prior contextual information. A corresponding drawback however lies in the need for many learning episodes which severely restricts the applicability of such approach in real-world time- and energy-constrained missions. Here, we propose a model-aided deep Q-learning approach that, in contrast to previous work, considerably reduces the need for extensive training data samples, while still achieving the overarching goal of DRL, i.e to guide a battery-limited UAV on an efficient data harvesting trajectory, without prior knowledge of wireless channel characteristics and limited knowledge of wireless node locations. The key idea consists in using a small subset of nodes as anchors (i.e. with known location) and learning a model of the propagation environment while implicitly estimating the positions of regular nodes. Interaction with the model allows us to train a deep Q-network (DQN) to approximate the optimal UAV control policy. We show that in comparison with standard DRL approaches, the proposed model-aided approach requires at least one order of magnitude less training data samples to reach identical data collection performance, hence offering a first step towards making DRL a viable solution to the problem.
\end{abstract}





\section{Introduction}\label{sec:Intro}

Rapid innovation in producing low-cost commercial unmanned aerial vehicles (UAVs) has opened up numerous opportunities in the UAV market which is projected to reach 63.6 USD billion by 2025 \cite{Wood2019}. One key application scenario is the future Internet of Things (IoT), in which harvesting data from wireless nodes that are spread out over wide areas far away from base stations (BSs) generally requires higher transmission power to communicate the information, reducing the network’s operating duration by draining the sensor battery faster. A UAV that acts as a flying BS can describe a flight pattern that brings it in close range to the ground nodes, hence reducing battery consumption and increasing the energy efficiency of the data harvesting system. However, delivering this gain hinges on the availability of efficient methods to design a trajectory for the UAV, deciding when and where to collect data from ground nodes.

The popularity of deep reinforcement learning (DRL) in this context can be explained by the fact that full information about the scenario environment (e.g. IoT sensor positions) is not a prerequisite. Further reasons include the computational efficiency of DRL inference, as well as the inherent complexity of UAV path planning, which is in general non-convex and often NP-hard \cite{Zeng2019, Li2020}. However, one of the greatest obstacles to deploying DRL-based path planning to real-world autonomous UAVs is the prohibitively extensive training data required \cite{DulacArnold2019}, equivalent to thousands of training flights. In this work, we address this issue by proposing a so-called {\it model-aided DRL} approach that only requires a minimum of training data to control a UAV data harvester under a limited flight-time.


The training data demand of DRL methods for UAV path planning depends in large parts on the scenario complexity and the availability of prior information about the environment. On the one hand, works such as \cite{Bayerlein2018asilomar}, where a deep Q-network (DQN) is trained to control an energy-limited UAV BS, assume absolutely no prior knowledge of the environment, requiring large amounts of training even in a simple environment as the DRL agent has to deduce the scenario conditions purely by trial and error. On the other hand, near perfect state information in works such as \cite{Zhang2020}, where cooperative UAVs are tasked with collecting data from IoT devices in a relatively simple unobstructed environment, enables faster convergence and requires less training data. In this work, prior knowledge available to the UAV agent is in between the two extremes: while some reference IoT node positions are known (referred to as anchors), other node positions and the challenging wireless channel characteristics in a dense urban environment that causes alternation between \ac{los} and \ac{nlos} links, must be estimated.

In the context of sample-efficient RL, model-accelerated solutions have been proposed previously for a variety of applications. A method called \textit{imagination rollouts} to increase sample-efficiency for a continuous Q-learning variant has been suggested for simulated robotic tasks in \cite{Gu2016}. Their approach is based on using iteratively refitted time-varying linear models, in contrast to a neural network (NN) model that we propose here. Learning a NN model in the context of stochastic value gradient learning methods has been proposed in \cite{Heess2015}.

Other works in the area of RL trajectory optimization for UAV communications have suggested other ways of reducing training data demand. Li \textit{et al.} \cite{Li2020} proposed a DRL method for sum-rate maximization from moving users based on transfer learning to reduce training time. In \cite{HuChen2020}, the authors propose meta-learning on random user uplink access demands for distributed UAV BS control, reducing training time by around 50\% compared to standard RL. Another possibility as proposed in \cite{Bayerlein2020} is to directly generalize training over a range of likely scenario parameters. The DRL agent then requires no retraining when scenario parameters randomly change at the cost of longer initial training and the requirement for the change being observable for the agents.

To the best of our knowledge, this is the first work that proposes model-based acceleration of the training process in DRL UAV path planning and also the first one that suggests the use of anchor nodes. Our contributions are as follows:
\begin{itemize}
\item We propose a novel model-aided DRL UAV path planning algorithm for data collection from IoT devices that requires a minimum of expensive real-world training data.
\item By introducing a device localization algorithm that exploits a limited number of reference device positions and a city 3D map, we show that our proposed method offers fast convergence even under uncertainty about device positions and without prior knowledge of the challenging radio channel conditions in a dense urban environment.
\item We compare our model-aided approach to the baselines of standard DRL without any prior information as well as map-based full knowledge DRL and show, that our approach achieves a reduction in training data demand of at least one order of magnitude with identical data collection performance.
\end{itemize}

\section{System Model and Problem Formulation}\label{sec:SysModel}
We consider a wireless communication system where a UAV-mounted flying BS is serving $K$ static ground level nodes (IoT sensors) in an urban area. The $k$-th ground node, $k\in\left[1,K\right]$, is located at ${\bf u}_{k}\in\mathbb{R}^{2}$. The ground nodes are split into two groups: nodes with known locations ${\bf u}_{k}, k\in \mathcal{U}_{\text{known}}$, and nodes with unknown locations ${\bf u}_{k}, k\in \mathcal{U}_{\text{unknown}}$.

The UAV mission lasts for a maximum duration of $T$ during which the UAV follows a trajectory with a constant velocity to maximize the amount of data collected from ground nodes. For the ease of exposition, we assume that the time period $T$ is discretized into $N$ equal time slots. The UAV position at time step $n$ is denoted by ${\bf v}_n=[x_n,y_n,h]^{\Tr}\in\mathbb{R}^{3}$,
where $h$ represents the altitude of the drone. We also assume that the drone is equipped with a GPS receiver, hence the coordinates ${\bf v}_n,~n\in[1,N]$ are known.

\subsection{UAV Model}\label{sec:UAVModel}
During the mission, the drone's position evolves as
\begin{equation}
{\bf v}_{n+1} = {\bf v}_n+ {\bf a}_n \: ,\, {\bf a}_n \in \mathcal{A}, 
\label{eq:UAV_model}
\end{equation}
where ${\bf a}_n$ is the UAV movement action, and $\mathcal{A}$ is the set of feasible actions for the UAV given by
\begin{equation}
    \mathcal{A} = \left\{ 
\underset{\text{hover}}{\underbrace{\left[\begin{array}{c}
0\\
0\\
0
\end{array}\right]}},
\underset{\text{right}}{\underbrace{\left[\begin{array}{c}
c\\
0\\
0
\end{array}\right]}},
\underset{\text{left}}{\underbrace{\left[\begin{array}{c}
-c\\
0\\
0
\end{array}\right]}},
\underset{\text{up}}{\underbrace{\left[\begin{array}{c}
0\\
c\\
0
\end{array}\right]}},
\underset{\text{down}}{\underbrace{\left[\begin{array}{c}
0\\
-c\\
0
\end{array}\right]}}
\right\},
 \label{eq:action_space}
\end{equation}
where $c$ is the distance that the UAV travels within each time step. Moreover, the UAV is subject to a limited flying time depending on its battery budget. We indicate the remaining battery budget of the UAV at $n$-th time step by $b_n \in \mathbb{R}$ and it changes according to 
\noindent 
\begin{equation}
b_{n+1} = \begin{cases}
b_n - 0.5, & {\bf a}_n=\text{hover}\\
b_n - 1, & \text{otherwise.}
\end{cases}
\label{eq:Battery_model}
\end{equation}

\subsection{Channel Model}
We now describe the radio channel model that is used for computing the channel gains between the UAV and the ground nodes. Note that the channel model and the channel parameters are unknown to the UAV. Classically, the channel gain between two radio nodes which are separated by distance $d$ meters in dB is modeled as \cite{ChenYanGes}
\begin{equation}
g_{z}=\ss{}_{z}-10\,\alpha_{z}\log_{10}\left(d\right)+\eta_{z},
\label{eq:CH_Model_dB}
\end{equation}
where $\alpha_{z}$ is the path loss exponent, $\ss_{z}$ is the log of average channel gain at the reference point $d=1\si{m}$, $\eta_{z}$ stands for the shadowing component that is modeled as a Gaussian random variable with $\mathcal{N}(0,\sigma_{z}^{2})$. $z\in\left\{ \ac{los},\text{NLoS}\right\}$ emphasizes the strong dependence of the propagation parameters on the \ac{los} or \ac{nlos} condition. Note that \eqref{eq:CH_Model_dB} represents the logarithm of the channel gain which is averaged over the small scale fading of unit variance.


\subsection{Problem Formulation} \label{sec:Trajectory_optimization}
We are seeking to find an optimal trajectory for the UAV to maximize the overall collected data from all ground nodes within the UAV mission time. We assume that the ground nodes are served by the drone in a time-division multiple access (TDMA) manner where all ground nodes have an equal communication time access to the channel and are served sequentially. The ground node scheduling is performed automatically by the UAV and is not part of the optimization problem. Hence, for the node $k$ at time step $n$, the maximum throughput is given by
\begin{equation}\label{eq:ch_capacity}
C_{k, n}=\frac{1}{K}\log_{2}\left(1+\frac{P 10^{0.1 g_{n,k}}}{\sigma^{2}}\right),
\end{equation}
where $K$ is the number of ground nodes and $\frac{1}{K}$ is the normalization factor capturing the TDMA channel sharing effect, $g_{n,k}$ is the channel gain between the $k$-th
node and the UAV at time step $n$, $P$ denotes the up-link transmission
power of the ground node, and the additive white Gaussian noise power
at the receiver is denoted by $\sigma^{2}$. 

We can now formulate the problem of maximum data collection by taking into account the UAV mobility constraints as follows
\noindent
\begin{subequations}\label{eq:GeneralOptProb}
\begin{align}
\max_{{\bf{ a}}_n}\  & \,\sum_{k\in [1, K]}\sum_{n\in [1, N]} C_{k, n}\label{eq:total_collected_data}\\
\mbox{s.t.\ } & \eqref{eq:UAV_model}, \, \eqref{eq:Battery_model}\\
 & {\bf v}_1={\bf v}_{\text{I}}, {\bf v}_N={\bf v}_{\text{F}}\label{eq:start_end_pos_const}\\
 & b_N \ge 0,\label{eq:battery_const}
\end{align}
\label{eq:general_trj_problem}
\end{subequations}
where \eqref{eq:total_collected_data} is the total collected data from all nodes during the mission, ${\bf v}_{\text{I}}, {\bf v}_{\text{F}}$ are, respectively, the starting and the final points of the trajectory, and \eqref{eq:battery_const} guarantees that there is enough battery power to reach the terminal point. This problem is challenging to solve, since the objective function \eqref{eq:total_collected_data} is highly non-convex and also the channel model and some of the ground nodes locations are not available at the UAV side. 


\section{Markov Decision Process and Q-Learning} \label{sec:standard_Q_RL}
To solve problem \eqref{eq:general_trj_problem}, we first reformulate it as a Markov decision process (MDP) which is defined by a $4$-tuple $(\mathcal{S}, \mathcal{A}, P_{\bf{a}}, R_{\bf{a}})$ with state space $\mathcal{S}$, actions space $\mathcal{A}$, the state transition probability function $P_{\bf{a}}$ giving the probability that action ${\bf{a}}$ in state $s$ at time step $n$ will lead to state $s'$ in the next time step, and the reward function $R_{\bf{a}}(s, s')$ which yields the immediate reward received after transitioning from state $s$ to state $s'$ by taking action ${\bf{a}}$. In our problem, each state comprises two elements which is given by $s_n=({\bf{v}}_n, b_n)$, and the action space $\mathcal{A}$ is defined in accordance with \eqref{eq:action_space}. 



The reward function consists of two components
\noindent 
\begin{equation}
    r_n = \sum_{k\in [1, K]} C_{k, n} - \lambda_n, \label{eq:instantaneous_reward}
\end{equation}
where $r_n \triangleq R_{\bf{a}}(s, s')$. The first term in \eqref{eq:instantaneous_reward} is the instantaneous collected data from all nodes at the $n$-th time step, and $\lambda_n$ is a penalty imposed by the safety controller that guarantees the UAV will reach the terminal point ${\bf{v}}_{\text{F}}$. Specifically, the safety controller at each time step computes the shortest trajectory (a minimum set of actions) and the minimum required power for getting to the destination point from the current UAV location, then based on these values it declines or accepts the current action ${\bf{a}}_n$ chosen by the UAV. If action ${\bf{a}}_n$ is rejected, a penalty term will be added to the reward function. The shortest trajectory and the minimum required power computed at the $n$-th time step by the safety controller are denoted by $\mathcal{A}^{sc}_{n}$ and $b^{sc}_{n}$, respectively. Thus, the safety penalty $\lambda_n$ is given by
\begin{equation}
    \lambda_n = \begin{cases}
\lambda, & b_n \leq b^{sc}_{n}\\
0, & \text{else}.
\end{cases}
\end{equation}
The action chosen by the UAV at each time step is checked and modified (if necessary) by the safety controller as follows
\noindent 
\begin{equation}\label{eq:safety_controller}
    {\bf{a}}_n = \begin{cases}
{\bf{a}}_{n, 1}^{sc}, & b_n \leq b^{sc}_{n} \wedge {\bf{a}}_n \notin \mathcal{A}^{sc}_{n}\\
{\bf{a}}_n, & \text{else},
\end{cases}
\end{equation}
where ${\bf{a}}_{n, 1}^{sc}$ is the first element of $\mathcal{A}^{sc}_{n}$.

To solve the MDP, we employ the popular Q-learning algorithm, a model-free RL technique, that enables us to directly compare our proposed method to the state-of-the-art from the literature. Note that, our aim is to reduce the real-world training data samples of Q-learning by model-aided acceleration with an \textit{external} model that simulates the environment. Accordingly, the Q-learning algorithm is unchanged and follows the standard cycle of interaction between agent and environment to iteratively learn a policy $\pi(s)$ that tells the agent how to select actions given a certain state.

Q-learning relies on iteratively improving the state-action value function $Q^{\pi}$, a.k.a. Q-function. The Q-function represents an expectation of the total future reward when taking action ${\bf{a}}$ in state $s$ and then following policy $\pi$. It is given by
\noindent 
\begin{equation}
  Q^{\pi}(s, {\bf{a}}) = \Exp_{\pi}\left[\sum_{m=n}^{N}\gamma^{m-n} r_n| s_n=s, {\bf{a}}_n={\bf{a}}\right],
\end{equation}
with discount factor $\gamma^{m-n}\in [0, 1]$ striking a balance between the importance of immediate and future rewards.

In large state-action spaces, the Q-function is commonly approximated by a Deep Q-network (DQN) with the neural network parameters $\theta$ \cite{Bayerlein2018asilomar}. When training the DQN $Q^{\pi}(s, {\bf{a}}; \theta)$, instability can occur. Experience replay, where experience tuples $(s_n, {\bf{a}}_n, r_n, s_{n+1} )$ are stored to be reused for training by the agent, and a separate target network with parameters $\hat{\theta}$ have become standard techniques to mitigate the risk of training instability \cite{Mnih2015}. Accordingly, the loss function at each time step to train the DQN is given by
\begin{equation} \label{eq:DQL-target-loss}
    \ell(\theta)=\Exp\left[\left(r + \gamma \max_{ {\bf{a}}} Q^{\pi}(s^\prime, {\bf{a}}; \hat{\theta} ) - Q^{\pi}(s, {\bf{a}}; \theta)\right)^2\right],
\end{equation}
where $Q^{\pi}(s_{n+1}, {\bf{a}}; \hat{\theta})$ is the target network and with time index $n$ omitted and $s_{n+1}$ abbreviated to $s^\prime$ for brevity.

\section{Model-aided Deep Q-Learning} 
Employing standard deep Q-learning is often not practical due to the tremendous amount of training data points required and the cost associated with obtaining these data points, i.e. through real-world UAV experiments. To ameliorate this problem, we propose an algorithm where the agent learns an environment model continuously while collecting real-world measurements. This model is then used by the agent to simulate experiments and supplement the real-world data.


More specifically in our scenario, the next state $s_{n+1}$ given the current state $s_n$ and action ${\bf{a}}_n$ can be computed from \eqref{eq:UAV_model}, \eqref{eq:Battery_model}. The reward function \eqref{eq:instantaneous_reward} consists of two parts: the safety penalty, which is known from \eqref{eq:safety_controller}, and the instantaneous collected data from the IoT node devices. Therefore we only need to estimate the instantaneous collected data from devices which according to \eqref{eq:ch_capacity}, \eqref{eq:CH_Model_dB}, is a function of ground node locations and the radio channel model. Hence, the approximation of the reward function boils down to ground node localization and radio channel learning from collected radio measurements.

The problem of simultaneous wireless node localization and channel learning has been studied in previously \cite{esrafilian20203d}. In this section, we propose a new approach of model-free node localization by leveraging the 3D map of the environment. Akin to \cite{esrafilian20203d}, a LoS/NLoS segmented radio channel is assumed. However, in contrast to \cite{esrafilian20203d}, our goal here is to estimate the radio channel using a model-free method while localizing the ground nodes. To learn the radio channel, we use a neural network (NN). This network is utilised along with a particle swarm optimization (PSO) technique and a 3D map of the city to localize the wireless nodes with unknown positions.


\subsection{Simultaneous Node Localization and Channel Learning}

We assume the UAV follows an arbitrary trajectory denoted by $\chi = \left\{{\bf{v}}_n, n\in [1,N]\right\}$ for collecting received signal strength (RSS) measurements, where ${\bf{v}}_n$ represents the UAV's position in the $n$-th time interval. We also assume that the UAV collects radio measurements form all $K$ nodes at each location. Let $g_{n,k}$ represent the RSS measurements (in dB scale) obtained from
the $k$-th node by the UAV in the $n$-th interval. Assuming a LoS/NLoS segmented pathloss model that is suitable for air-to-ground channels in urban environments with buildings \cite{ChenYanGes}, we have
\begin{equation}\label{eq:chGainModel}
g_{n,k} {=}
\begin{cases} 
    \psi_{{\boldsymbol{\theta}}} (d_{n,k}, \phi_{n,k},w_{n,k}{=}1) + \eta_{n,k,\ac{los}}       & \small{\text{if} \text{ LoS}}\\
    \psi_{{\boldsymbol{\theta}}} (d_{n,k}, \phi_{n,k},w_{n,k}{=}0) + \eta_{n,k,\ac{nlos}}       & \small{\text{if} \text{ NLoS}},
   \end{cases}
\end{equation} 
where $d_{n,k} = \| {\bf{u}_k}-{\bf{v}_n} \|$, and $\phi_{n,k}=\arcsin(\frac{\bar{d}_{n,k}}{d_{n,k}})$ is the elevation angle between the UAV at time step $n$ and node $k$ with $\bar{d}_{n,k}$ representing the ground distance between the ground node and the UAV. $\omega_{n,k} \in \{0,1\}$ is the classification binary variable (yet unknown) indicating whether a measurement falls into the LoS or NLoS category. The function $\psi_{{\boldsymbol{\theta}}} (.)$ is the channel model parameterized by ${\boldsymbol{\theta}}$. Note that, neither function $\psi (.)$ nor parameters ${\boldsymbol{\theta}}$ are known and need to be estimated. $\eta_{n,k,z}$ stands for the shadowing effect with zero-mean normal distribution with known variance $\sigma_{z}^{2}$. The probability distribution of a single measurement in \eqref{eq:chGainModel} is modeled as
\begin{equation} \label{eq:mixdit}
    p(g_{n,k}) = (f_{n,k,\text{LoS}})^{w_{n,k}} (f_{n,k,\text{NLoS}})^{(1-w_{n,k})},
\end{equation}
where $f_{n,k,z}$ has a Gaussian distribution with $\mathcal{N}(\psi_{{\boldsymbol{\theta}}} (d_{n,k}, \phi_{n,k},w_{n,k}),\sigma_{z}^{2})$.

Assuming that collected measurements conditioned on the channel and node positions are independent and identically distributed (i.i.d) \cite{ChenYanGes}, using \eqref{eq:mixdit}, the negative log-likelihood of measurements leads to 
\noindent 
\begin{equation}
\begin{aligned}\label{eq:localization_liklihood}
\mathcal{L} =&  \log\left(\frac{\sigma_{\ac{los}}^2}{\sigma_{\ac{nlos}}^2}\right)\sum_{k=1}^{K}\sum_{n=1}^{N} \omega_{n,k}+ \\ &\sum_{k=1}^{K}\sum_{n=1}^{N}
  \frac{\omega_{n,k}}{\sigma_{\ac{los}}^2}\left |g_{n,k}{-}\psi_{{\boldsymbol{\theta}}} (d_{n,k}, \phi_{n,k},w_{n,k})\right|^2 + \\
 & \sum_{k=1}^{K}\sum_{n=1}^{N} \frac{(1-\omega_{n,k})}{\sigma_{\ac{nlos}}^2}  \left | g_{n,k}{-}\psi_{{\boldsymbol{\theta}}} (d_{n,k}, \phi_{n,k},w_{n,k})\right|^2.
\end{aligned}
\end{equation}
The estimate of $\psi(.), {\boldsymbol{\theta}},$ and $\bf{u}_k$ can then be obtained by solving
\begin{subequations}
\begin{align}
 \begin{split}
           \min_{\substack{\omega_{n,k},\,{{\bf{u}}}_k,\forall n,\forall k\\ {\psi(.), {\boldsymbol{\theta}}}}
} & \quad \mathcal{L} 
 \end{split}\\
   \begin{split}
          \text{s.t.}&\quad \omega_{n,k} \in \{0,1\}, \forall n, \forall k.
 \end{split}
\end{align}\label{eq:Localization_Opt_Org}
\end{subequations}
The binary variables $\omega_{n,k}$ in objective function \eqref{eq:localization_liklihood}, and the fact that $\psi(.)$ is not explicitly known and is a nonlinear function of node locations, make problem \eqref{eq:Localization_Opt_Org} challenging to solve since it is a joint classification, channel learning and node localization problem. To tackle this difficulty, we split \eqref{eq:Localization_Opt_Org} into two sub-problems of learning the channel and localizing nodes. We also leverage the 3D map of the city for the measurements classification which will be discussed next.

\subsubsection{Radio Channel Learning} \label{sec:cahnnel_learning}
Our aim is to learn the radio channel using collected radio measurements from the IoT nodes with known location (anchor nodes). Since the characteristic of the radio channel is independent of the node location and only affected by the structure of the city and the blocking objects in the environment, learning the radio channel from the nodes with known location can provide a good approximation of the radio channel. The measurements are classified by leveraging the 3D map of the city, since for a node with known location the classification variables $\omega_{n,k}$ can be directly inferred from a trivial geometry argument: for a given UAV position, the node is considered in LoS to the UAV if the straight line passing through the UAV's and the node position lies higher than any buildings in between. Having classified the measurements, we use a neural network with parameters ${{\boldsymbol{\theta}}}$ as an approximation of $\psi_{{{\boldsymbol{\theta}}}}(.)$. The neural network accepts an input vector $[d_{n,k}, \phi_{n,k},w_{n,k}]^{\text{T}}$ and returns an estimate of the channel gain $\hat{g}_{n,k}$. Therefore, problem \eqref{eq:Localization_Opt_Org} just by considering the anchor nodes can be rewritten as follows
\begin{equation}
\begin{aligned}
             \min_{\substack{{{\boldsymbol{\theta}}}\\ k\in \mathcal{U}_{\text{known}}, \forall n}
} & \quad \mathcal{L}.
\end{aligned} \label{eq:cahnnel_learning}
\end{equation}
This optimization is a standard problem in machine learning and can be solved using any gradient-based optimizer. The parameters obtained by solving \eqref{eq:cahnnel_learning} are denoted by ${\boldsymbol{\theta}}^*$.

\subsubsection{Node Localization} \label{sec:localization}
Having learned the radio channel, we continue to localize the unknown nodes. The optimization problem \eqref{eq:Localization_Opt_Org} for the set of unknown nodes and utilizing the learned radio channel can be reformulated as follows:
\begin{subequations}
\begin{align}
  \begin{split}
           \min_{\substack{\omega_{n,k},\,{{\bf{u}}}_k,\forall n\\ {k \in \mathcal{U}_{\text{unknown}}}}
} & \quad \mathcal{L}^* 
 \end{split}\\
   \begin{split}
          \text{s.t.}&\quad \omega_{n,k} \in \{0,1\}, k \in \mathcal{U}_{\text{unknown}}, \forall n,
 \end{split}
\end{align}\label{eq:Localization_unkown_users}
\end{subequations}
where $\mathcal{L}^* $ is obtained by substituting the learned channel model $\psi_{{{\boldsymbol{\theta}}}^*}(.)$ in \eqref{eq:localization_liklihood}. The binary random variables $\omega_{n,k}$, and the non-linear and non-convex objective function $\mathcal{L}^*$ make problem \eqref{eq:Localization_unkown_users} hard to solve. We use the PSO algorithm which is suitable for solving various non-convex and non-linear optimization problems. PSO is a population-based optimization technique that tries to find the solution to an optimization problem by iteratively trying to improve a candidate solution with regard to a given measure of quality (or objective function). The algorithm is initialized with a population of random solutions, called particles, and a search for the optimal solution is performed by iteratively updating each particle's velocity and position based on a simple mathematical formula (for more details on PSO see \cite{KenEbe}). As will be clear later, the PSO algorithm is enhanced to exploit the side information stemming from the 3D map of the environment which improves the performance of node localization and reduce the complexity of solving $\eqref{eq:Localization_unkown_users}$, since the binary variable $\omega_{n,k}$ can be obtained directly from the 3D map \cite{esrafilian20203d}.

For ease of exposition, we first solve $\eqref{eq:Localization_unkown_users}$ by assuming only one unknown node. Then we will generalize our proposed solution to the multi-node case. To apply the PSO algorithm, we define each particle to have the following form
\begin{equation}
    {\bf{c}}_j = [ x_j, y_j]^{\text{T}} \in \mathbb{R}^2, j \in [1, C],
\end{equation}
where $C$ is the number of particles and each particle is an instance of the possible node location in the city. Therefore, by treating each particle as a potential  candidate for the node location, the negative log-likelihood $\eqref{eq:localization_liklihood}$ for a given particle can be rewritten as follows
\noindent 
\begin{equation}\small
\begin{aligned} 
\mathcal{L}^*({\bf{c}}_{j}^{(i)}&) = \log\left(\frac{\sigma^2_{\ac{los}}}{\sigma^2_{\ac{nlos}}}\right) \left |\mathcal{M}_{\ac{los},1,j}\right| + \\ & \small{\sum_{z\in \{\ac{los},\ac{nlos}\}} \, \sum_{n\in \mathcal{M}_{z,1,j}}} \frac{1}{\sigma^{2}_{z}}
 \left |g_{n,1}{-}\psi_{{\boldsymbol{\theta}}^*} (d_{n,k}, \phi_{n,k},z)\right|^2  ,\label{eq:SSE_PSO_single_Ue}
\end{aligned}
\end{equation}

\noindent where ${\bf{c}}_{j}^{(i)}$ is the $j$-th particle at the $i$-th iteration of the PSO algorithm, and $\mathcal{M}_{z,1,j}$ is a set of time indices of measurements collected from node 1 which are in segment $z$ by assuming that the location of node 1 is the same as particle $j$. To form $\mathcal{M}_{z,1,j}$, a 3D map of the city is utilized. For example, measurement $g_{n,1}$ is considered LoS, if the straight line passing through ${\bf{c}}_{j}^{(i)}$ and the drone location ${\bf{v}}_n$ lies higher than any buildings in between. Therefore, the best particle minimizing \eqref{eq:SSE_PSO_single_Ue} can be obtained from solving the optimization
\noindent 
\begin{equation}
    j^* := \arg \min_{j\in [1,C]} {\mathcal{L}^{*}({\bf{c}}_{j}^{(i)})}, \label{eq:best_particle}
\end{equation}
where $j^* $ is the index of the best particle which minimizes the objective function in \eqref{eq:best_particle}. In the next iteration of the PSO algorithm, the position and the velocity of particles are updated and the algorithm repeats for $\tau$ iterations. The best particle position in the last iteration is considered as the estimate of the node location.

Note that for the multi-node case, without loss of optimality, the problem can be transformed to the multi single-node localization problem and then each problem can be solved individually. This stems from the fact that the radio channel is learned beforehand and is assumed to have the same characteristics for all the UAV-node links (the radio channel characteristics is assumed to be independent of the node locations).

\subsection{Algorithm}





The proposed Algorithm \ref{alg:Model_Aided_DQL} iterates between three phases: 1) the agent uses a policy obtained from its Q-network in the real world to collect RSS measurements from ground nodes. 2) The collected measurements are used to learn the radio channel and localize the unknown nodes as described in Sections \ref{sec:cahnnel_learning} and \ref{sec:localization}, respectively. 3) The agent performs a new set of experiments in the simulated environment under the learned radio channel model and the estimation of the node locations to train the Q-network. Then, the agent repeats the first phase of the algorithm by generating a new policy using the trained Q-network and the procedure continues until convergence of the Q-network training. 


The experience replay buffers for real world and simulated world experiments are denoted $\mathcal{B}$ and $\Tilde{\mathcal{B}}$, respectively. A new episode in phase 1 and phase 3 starts by resetting the time index, the initial UAV position and the battery budget (lines 7 and 22). To train the Q-network, an $\epsilon$-greedy exploration technique is used (line 36) with decay constant $\kappa$. $\beta$ is the learning rate for primary network parameters $\theta$. Target network parameters are updated every $N_{target}$ episodes. In phase 3, the algorithm performs $I$ sets of experiments in the simulated world, and the whole algorithm terminates after carrying out $E_{max}$ real-world experiments.

\begin{algorithm}
    \begin{algorithmic}[1]
    \fontsize{9}{9.8}\selectfont
    \STATE Initialize replay buffer $(\mathcal{B}), (\Tilde{\mathcal{B}})$
    \STATE Initialize Q-network and target network parameters 
    \STATE Initialize $t=0$
    \FOR{$e=0$ to $E_{max}$}
    \STATE $t = t + 1$
    \STATE {\bf{\text{1) }Real-world experiment:}}
        \STATE Initialize $s_0 = ({\bf{v}}_{\text{I}}, b_{max}), \,n=0$
        \WHILE {$b_n \ge 0$}
            \STATE $
                {\bf{a}}_n = \arg \max_{{\bf{a}}} Q^{\pi}(s_{n}, {\bf{a}}, \theta)$
            \STATE Validate ${\bf{a}}_n$ using the safety controller \eqref{eq:safety_controller}
            \STATE Observe $r_n, s_{n+1}, \gamma_{1, n}, \cdots, \gamma_{K, n} $
            \STATE Store $(s_n, {\bf{a}}_n, r_n, s_{n+1} )$ on $(\mathcal{B})$
            \STATE Memorize $({\bf{v}}_n, \gamma_{1, n}, \cdots, \gamma_{K, n})$ 
            \STATE $n=n+1$
        \ENDWHILE
        \STATE {\bf{\text{2) }Learning the environment:}}
        \STATE Learn the radio channel as described in Section \ref{sec:cahnnel_learning}
        \STATE Localize unknown nodes as described in Section \ref{sec:localization}
        \STATE {\bf{\text{3) }Simulated-world experiment:}}
        \FOR{$i=0$ to $I$}
        \STATE $t = t + 1$
        \STATE Initialize $\Tilde{s}_0 = ({\bf{v}}_{\text{I}}, b_{max}), \,n=0$
        \WHILE {$b_n \ge 0$}
            \STATE $
                \Tilde{{\bf{a}}}_n = \begin{cases}
                \text{randomly select from}\, \mathcal{A} & \text{with probability} \, \epsilon\\
            \arg \max_{{\bf{a}}} Q^{\pi}(\Tilde{s}_{n}, {\bf{a}}, \theta) & \text{else}
            \end{cases}
            $
            \STATE Validate $\Tilde{{\bf{a}}}_n$ using the safety controller \eqref{eq:safety_controller}
            \STATE Compute $\Tilde{r}_n$ from \eqref{eq:instantaneous_reward}, and $\Tilde{s}_{n+1}$ from \eqref{eq:UAV_model}, \eqref{eq:Battery_model}
            \STATE store $(\Tilde{s}_n, \Tilde{{\bf{a}}}_n, \Tilde{r}_n, \Tilde{s}_{n+1} )$ on $\Tilde{\mathcal{B}}$
            \FOR{$m=0$ to $M$}
                \STATE Sample $(s_m, {\bf{a}}_m, r_m, s_{m+1} )$ uniformly from $\{\mathcal{B} \cup \Tilde{\mathcal{B}}\}$
                \STATE $
                y_m = \begin{cases}
                 r_m & \text{if terminal}\\
            r_m + \gamma \max_{{\bf{a}}} Q^{\pi}(s_{m+1}, {\bf{a}}, \hat{\theta}) & \text{else}
            \end{cases}
            $
            \STATE $\ell_m(\theta) =  \Exp\left[\left( y_m - Q^{\pi}(s_{m}, {\bf{a}}_m, \theta) \right)^2 \right]$
            \ENDFOR
            \STATE $\theta = \theta  + \beta \frac{1}{M} \nabla_{\theta} \sum_{m=0}^M \ell_m(\theta)$
            \STATE $n=n+1$
        \ENDWHILE
        \STATE $\epsilon = \epsilon_{final} + (\epsilon_{start} - \epsilon_{final})\exp(-\kappa t)$
        \STATE $\bold{if}$ {$(t \mod N_{target} = 0)$} $\bold{then}$ $\hat{\theta} = \theta$
    \ENDFOR
    \ENDFOR
    \end{algorithmic}
    \caption{Model-aided deep Q-learning trajectory design}
    \label{alg:Model_Aided_DQL}
\end{algorithm}


\section{Numerical Results}\label{sec:simulations}
\begin{figure}[ht]
\begin{centering}
\includegraphics[width=0.83\columnwidth]{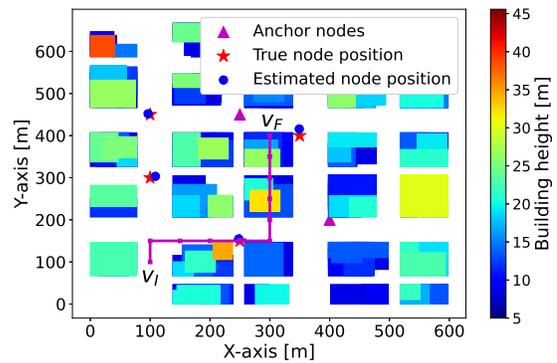}
\par\end{centering}
\caption{Trajectory obtained in the final episode of Algorithm 1 as well as the estimates of unknown node locations.
\label{fig:model_aided_trj_localiztion}}
\vspace{-6pt}
\end{figure}

We consider a dense urban city neighborhood comprising buildings and regular streets as shown in Fig. \ref{fig:model_aided_trj_localiztion}. The height of the buildings is Rayleigh distributed in the range of 5 to \SI{40}{m} and the true propagation parameters are chosen similar to \cite{Bayerlein2020}. The UAV collects radio measurements from the ground nodes every \SI{5}{m} and we assume that the altitude of the UAV is fixed to \SI{60}{m} during the course of its trajectory. The mission time of each episode is fixed to $N=20$ time steps with a fixed step size of $c=\SI{50}{m}$. 
We assume there are six ground nodes. Only the locations of anchor nodes ${\textbf{u}}_1$ and ${\textbf{u}}_2$ are known to the UAV in advance.
The UAV starts from ${\bf v}_{\text{I}}=[100, 100, 60]^{\text{T}}$ and needs to reach the destination point ${\bf v}_{\text{F}}=[300, 400, 60]^{\text{T}}$ by the end of the mission. To learn the channel, we use a NN with two hidden layers where the first layer has $60$ neurons with $tanh$ activation function, and the second layer $30$ neurons with $relu$ activation function. The Q-network comprises 2 hidden layers each with $120$ neurons and $relu$ activation function.

\begin{figure}[ht]
\begin{centering}
\includegraphics[width=0.65\columnwidth]{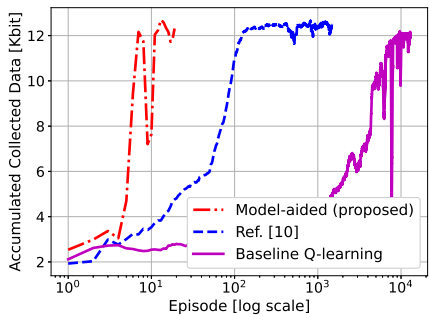}
\par\end{centering}
\caption{
Comparison of different algorithms, showing accumulated collected data versus training episodes.
\label{fig:algorithms_perf_comparison}}
\end{figure}

In Fig. \ref{fig:algorithms_perf_comparison}, we compare the performance of the baseline Q-learning algorithm as explained in Section \ref{sec:standard_Q_RL} and akin to \cite{Bayerlein2018asilomar}, with the proposed model-aided Q-learning algorithm. Moreover, we show the result of an algorithm similar to \cite{Bayerlein2020}, where the mixed-radio map of the nodes is embedded in the state vector. To compute the mixed-radio map, the individual radio maps of all nodes are combined. Individual radio maps are computed using the 3D map of the city and assuming perfect knowledge node positions and the radio channel. The model-aided algorithm outperforms the other approaches since it merely requires 10 real-world experiment episodes to converge to the same performance level as other algorithms. The algorithm introduced in \cite{Bayerlein2020} is superior to the baseline since it uses more information, i.e. the map and perfect knowledge of node positions and the radio channel model.


Fig. \ref{fig:model_aided_trj_localiztion} shows the final trajectory after convergence. The UAV starts flying towards the closest node and hovers above for several time steps in order to maximize the amount of collected data, and then reaches the destination ${\bf v}_{\text{F}}$. Moreover, the estimate of unknown node locations obtained at the last episode of the training phase of Algorithm \ref{alg:Model_Aided_DQL} are shown and confirmed to be very close the true positions.


\section{Conclusion}\label{sec:conclusion}
We have introduced a novel model-accelerated DRL path planning algorithm for UAV data collection from distributed IoT nodes with only partial knowledge of the nodes' locations. In comparison to two standard deep Q-learning algorithms, using either full or no knowledge of sensor node locations, we have demonstrated that the model-aided approach requires at least one order of magnitude less training data samples to reach the same data collection performance. 


\section{Acknowledgments}
This work was partially funded via the HUAWEI
France supported Chair on Future Wireless Networks at EURECOM and by the
CARNOT Institute Télécom $\&$ Société numérique, under project Robot4IoT.

\bibliographystyle{IEEEtran}
\bibliography{literature.bib}

\end{document}